\documentclass[12pt,english]{article}
\usepackage[a4paper, portrait,
	left=0.75in,
	right=0.75in,
	top=1in,
	bottom=0.5in,
	footskip=.25in]{geometry}
\usepackage[english]{babel}
\usepackage[utf8]{inputenc}
\usepackage[T1]{fontenc}
\usepackage{helvet,etoolbox,graphicx,titlesec}
\usepackage{caption,subcaption}
\usepackage[affil-it]{authblk} 
\usepackage{etoolbox}
\usepackage{lmodern}

\usepackage{mathrsfs}
\usepackage{amsfonts}
\usepackage{amsmath} 
\usepackage{amssymb} 

\usepackage{titlesec}

\usepackage{mathrsfs}

\usepackage{graphicx}
\usepackage{dcolumn}
\usepackage{bm}

\usepackage{amsthm}
\usepackage{xspace}
\usepackage{xfrac}
\usepackage[mathcal]{euscript}
\usepackage{tikz}
\usepackage{ifthen}

\usepackage{pgfplots}
\usepgfplotslibrary{patchplots,colormaps}
\pgfplotsset{width=7cm,compat=1.18,
colormap={mycolormap}{color=(black) color=(black!20!white)}}
\usepgfplotslibrary{fillbetween} 
\usepgflibrary{shadings}
\usetikzlibrary{backgrounds}

\usepackage{epsdice}
\usepackage{twemojis}

\usepackage{pifont}

\usepackage{tcolorbox}
\tcbuselibrary{theorems,skins,breakable}
\usepackage{float}
\usepackage{colortbl}

\usepackage{natbib}

\usepackage{orcidlink}

\usepackage{url}
\usepackage{hyperref}
\usepackage{color}
\definecolor{refcolor}{RGB}{160,35,0}
\definecolor{hrefcolor}{RGB}{0,35,190}
\hypersetup{
    colorlinks,
    citecolor=refcolor,
    filecolor=refcolor,
    linkcolor=hrefcolor,
    urlcolor=hrefcolor
}

\usepackage{booktabs}
\usepackage{bookmark}
\bookmarksetup{
  numbered, 
  open,
}

\definecolor{greenPsi}{rgb}{0.0, 0.375, 0.0}
\definecolor{blueStruct}{rgb}{0.0, 0.0, 1.0}
\definecolor{redStruct}{rgb}{1.0, 0.0, 0.0}

\newcommand{\boxRule}{0.15mm}

\newcommand{\boxIndent}{15pt}

\newcommand{\remColor}{green}
\newcommand{\quoteColor}{green!50!yellow}
\newcommand{\questColor}{red}
\newcommand{\figColor}{orange}
\newcommand{\tabColor}{green}
\newcommand{\controlColor}{yellow}
\newcommand{\abstractColor}{blue!80!cyan}

\newenvironment{frameEnv}[1]
	{\begin{tcolorbox}[breakable,enhanced,toprule at break=0pt,bottomrule at break=0pt,before skip balanced=0.3cm,boxrule=\boxRule,left=0.75mm,right=0.75mm,frame hidden,borderline north = {\boxRule}{0pt}{#1!50!black}, borderline south = {\boxRule}{0pt}{#1!50!black},arc=0mm,colframe=#1!50!black,colback=#1!10,before upper={\parindent\boxIndent}]}
	{\end{tcolorbox}}

\newenvironment{fquest}
	{\begin{frameEnv}{\questColor}}
	{\end{frameEnv}}
\newenvironment{fquote}
	{\begin{frameEnv}{\quoteColor}}
	{\end{frameEnv}}
\newenvironment{ffig}
	{\begin{frameEnv}{\figColor}}
	{\end{frameEnv}}
\newenvironment{ftab}
	{\begin{frameEnv}{\tabColor}}
	{\end{frameEnv}}
\newenvironment{fcontrol}
	{\begin{frameEnv}{\controlColor}}
	{\end{frameEnv}}

	\newenvironment{frameEnvMargin}[1]
	{\begin{tcolorbox}[breakable,enhanced,toprule at break=0pt,bottomrule at break=0pt,before skip balanced=0.3cm,boxrule=\boxRule,left=0.75mm,right=0.75mm,top=5mm,bottom=5mm,frame hidden, borderline north = {\boxRule}{0pt}{#1!50!black}, borderline south = {\boxRule}{0pt}{#1!50!black},arc=0mm,colframe=#1!50!black,colback=#1!10,before upper={\parindent\boxIndent}]}
	{\end{tcolorbox}}

\newenvironment{fabstract}
	{\begin{frameEnvMargin}{\abstractColor}\begin{abstract}}
	{\end{abstract}\end{frameEnvMargin}}

\theoremstyle{remark}

\numberwithin{proofStep}{theorem} 

\theoremstyle{definition}
\newtheorem{defCustom}{Definition of}
\renewcommand{\thedefCustom}{\arabic{definition}}
\makeatletter
\newcommand{\setdefCustomtag}[1]{
  \let\oldthedefCustom\thedefCustom
  \renewcommand{\thedefCustom}{#1}
  \g@addto@macro\enddefCustom{
    \global\let\thedefCustom\oldthedefCustom}
  }
\makeatother

\theoremstyle{definition}

\newtheorem{feat}{Feature}
\newtheorem{chcount}{Count}

\theoremstyle{remark}
\newtheorem{note}{Note}

\newtheorem{condition}{Condition}
\renewcommand{\thecondition}{\arabic{condition}}
\makeatletter
\newcommand{\setconditiontag}[1]{
  \let\oldthecondition\thecondition
  \renewcommand{\thecondition}{#1}
  \g@addto@macro\endcondition{
    \global\let\thecondition\oldthecondition}
  }
\makeatother

\theoremstyle{remark}

\makeatletter
\newcommand*{\radiobutton}{%
  \@ifstar{\@radiobutton0}{\@radiobutton1}%
}
\newcommand*{\@radiobutton}[1]{%
  \begin{tikzpicture}
    \pgfmathsetlengthmacro\radius{height("X")/2}
    \draw[radius=\radius] circle;
    \ifcase#1 \fill[radius=.6*\radius] circle;\fi
  \end{tikzpicture}%
}
\makeatother

\makeatletter
\newcommand*{\checkboxbutton}{%
  \@ifstar{\@checkboxbutton0}{\@checkboxbutton1}%
}
\newcommand*{\@checkboxbutton}[1]{%
\ifcase#1
$\text{\rlap{$\checkmark$}}\square$
\else
$\square$
\fi
}
\makeatother

\colorlet{disabled}{gray!50!black}

\newcommand{\orcid}[1]{\href{https://orcid.org/#1}{\textcolor[HTML]{A6CE39}{\aiOrcid}}}

\def\({\left(}
\def\){\right)}
\def\[{\left[}
\def\]{\right]}

\usepackage{authblk}

\newsavebox\affbox
\author{Cristi Stoica\ \orcidlink{0000-0002-2765-1562}}
\affil{Dept. of Theoretical Physics, NIPNE---HH, Bucharest, Romania.\\
Email: \textit{\color{cyan}\href{mailto:cristi.stoica@theory.nipne.ro}{cristi.stoica@theory.nipne.ro},  \href{mailto:holotronix@gmail.com}{holotronix@gmail.com}}}

\makeatletter
\newcommand*\@secondofsix[6]{#2}
\newcommand{\addtotitleformat}{%
  \@ifstar{\addtotitleformat@star}{\addtotitleformat@nostar}}
\newcommand\addtotitleformat@nostar[2]{%
  \PackageError{titlesec}{non starred form of \string\addtotitleformat\space not supported}{}}
\newcommand\addtotitleformat@star[2]{%
  \expandafter\expandafter\expandafter\expandafter
  \expandafter\expandafter\expandafter\def
  \expandafter\expandafter\expandafter\expandafter
  \expandafter\expandafter\expandafter\@currentsection@font
  \expandafter\expandafter\expandafter\expandafter
  \expandafter\expandafter\expandafter{%
    \expandafter\expandafter\expandafter\@secondofsix
       \csname ttlf@\expandafter\@gobble\string#1\endcsname}%
  \titleformat*{#1}{\@currentsection@font#2}%
}
\makeatother

\titlespacing\section{0pt}{20pt plus 6pt minus 4pt}{16pt plus 4pt minus 4pt}
\titlespacing\subsection{12pt}{12pt plus 6pt minus 4pt}{10pt plus 4pt minus 4pt}
\titlespacing\subsubsection{12pt}{12pt plus 6pt minus 4pt}{10pt plus 4pt minus 4pt}

\titleformat{\section}{\normalfont\fontsize{16}{24}\bfseries}{\thesection.}{1em}{}
\titleformat{\subsection}{\normalfont\fontsize{14}{20}\bfseries}{\thesubsection.}{1em}{}
\titleformat{\subsubsection}{\normalfont\fontsize{13}{18}\bfseries}{\thesubsubsection.}{1em}{}

\titleformat{\author}{\normalfont\fontsize{14}{20}\bfseries}{\thesection}{1em}{}

\makeatletter
\renewcommand{\thesubsection}{\arabic{section}.\arabic{subsection}}
\makeatother



\definecolor{titcolor}{RGB}{0,90,255}
\addtotitleformat*{\section}{\Large\sffamily\color{titcolor}}
\addtotitleformat*{\subsection}{\large\sffamily\color{titcolor}}
\addtotitleformat*{\subsubsection}{\large\sffamily\color{titcolor}}

\title{\color{titcolor}\textbf{Freedom in the Many-Worlds Interpretation}}

\date{\small\today} 

\begin{document}

\pagestyle{headings}	
\newpage
\setcounter{page}{1}
\renewcommand{\thepage}{\arabic{page}}

\maketitle

\begin{fabstract}
I analyze the possibility of free-will in the many-worlds interpretation (MWI), arguing for their compatibility. I use as a starting point Nicolas Gisin's ``The Multiverse Pandemic'' \citep{Gisin2022MultiversePandemic,Gisin2010LEpidemieDuMultivers}, 
in which he makes an interesting case that MWI is contradicted by our hard to deny free-will. The counts he raised are:

(1) MWI is deterministic, forcing choices on us,

(2) in MWI all our possible choices happen, and

(3) MWI limits creativity, because everything is entangled with everything else.

I argue that each of these features of MWI is in fact compatible with more freedom than it may seem. In particular, MWI allows compatibilist free-will, but also free-will very much like the libertarian free-will defined by Chisholm. I argue that the position that alternative choices exist as possibilities does not make sense from a physical point of view, but MWI offers a physical ground for alternatives.
\end{fabstract}


\maketitle

\section{Introduction}
\label{s:intro}

This article focuses on free-will only in relation with physics, especially with quantum mechanics, in particular the many-worlds interpretation. I will ignore the aspects of free-will related to legal and moral responsibility, but also numerous other related themes, including philosophy and religion and their rich histories. There are many others who analyzed and discussed all these much better than I could ever do, here is a selection \citep{sep-freewill,Dilman2013FreeWillAnHistoricalAndPhilosophicalIntroduction,Dennett2015ElbowRoomTheVarietiesOfFreeWillWorthWanting,Strawson1998FreeWillRoutledge,Balaguer2012FreeWillAsAnOpenScientificProblem,Kane1998TheSignificanceOfFreeWill,Kane2005AContemporaryIntroductionToFreeWill,Swinburne2013MindBrainAndFreeWill,Sapolsky2023DeterminedLifeWithoutFreeWill}.

I will use as a starting point for this discussion of free-will in the context of the many-worlds interpretation of quantum mechanics Nicolas Gisin's criticism \citep{Gisin2022MultiversePandemic,Gisin2010LEpidemieDuMultivers}, since it touches on important questions and it is well-written.

Perhaps the most important division of the views on free-will is based on the existence of indeterminism and its role.

\emph{Libertarian free-will} requires indeterminism, so that the free agent is able to make choices free from both external and internal constraints, including even the agent's own motives and tendencies \citep{Kane1998TheSignificanceOfFreeWill}. It is important that indeterminism is used in the process of decision-making, but not in a ``passive'' way, because indeterminism by itself doesn't guarantee free-will any more than basing your choices on tossing a coin or on using Vaidman's \emph{Quantum World Splitter} \citep{Vaidman-QuantumWorldSplitter} does. 
The agent should be the source of his choices or actions, it should initiate them, unaffected by prior events, and so that it could have made a different choice.
According to Chisholm \citep{Chisholm1964HumanFreedomAndTheSelf}:

\begin{fquote}
If we are responsible, and if what I have been trying to say is true, then we have a prerogative which some would attribute only to God: each of us, when we act, is a prime mover unmoved. In doing what we do, we cause certain events to happen, and nothing--or no one--causes us to cause those events to happen.
\end{fquote}

Here is an illustration of libertarian free-will, based on the graphical user interfaces used on computers. Alice is an agent that makes a choice between two options, having a coffee, or having a tea. She can choose to have a coffee,

\begin{fcontrol}
\begin{equation}
\label{eq:radiobutton-coffee}
\begin{aligned}
&\text{\radiobutton*}\text{ Alice chooses coffee.}\\
&\text{\radiobutton }\text{ Alice chooses tea.}\\
\end{aligned}
\end{equation}
\end{fcontrol}
\noindent or she can choose to have a tea
\begin{fcontrol}
\begin{equation}
\label{eq:radiobutton-tea}
\begin{aligned}
&\text{\radiobutton }\text{ Alice chooses coffee.}\\
&\text{\radiobutton*}\text{ Alice chooses tea.}\\
\end{aligned}
\end{equation}
\end{fcontrol}

By contrast, according to \emph{compatibilism}, not only free-will is compatible with determinism, but determinism is necessary for freedom, because it allows the agent to decide and act based on its own past history, desires, motives, and tendencies. For compatibilist free-will, we need at least ``effective determinism'', by which I mean that even if at the micro-physical scale determinism is not true, the decision-making is shielded from indeterminism so that it is, for all practical purposes, a deterministic process.
The agent is the source of her own actions, without needing to choose among alternatives. For example, if you want to drive a car, play piano, type some text, indeterminism, or even deterministic but chaotic causes, may affect your ability to control yourself and to exercise your freedom.
This doesn't necessarily require that determinism is valid down to the scale of micro-physics, but at least that it is effectively valid in the agent's decision-making.

Compatibilist free-will would determine Alice to have coffee, excluding tea,
\begin{fcontrol}
\begin{equation}
\label{eq:radiobutton-x-coffee}
\begin{aligned}
&\text{\radiobutton*}\text{ Alice chooses coffee.}\\
&\text{\textcolor{disabled}{\radiobutton}}\text{ \textcolor{disabled}{Alice chooses tea.}}\\
\end{aligned}
\end{equation}
\end{fcontrol}
\noindent or to have a tea, excluding coffee
\begin{fcontrol}
\begin{equation}
\label{eq:radiobutton-x-tea}
\begin{aligned}
&\text{\textcolor{disabled}{\radiobutton}}\text{ \textcolor{disabled}{Alice chooses coffee.}}\\
&\text{\radiobutton*}\text{ Alice chooses tea.}\\
\end{aligned}
\end{equation}
\end{fcontrol}

\noindent (I used the convention that the options excluded by determinism are ``disabled''.)

\emph{Hard incompatibilism} is the position that neither libertarian nor compatibilist types of free-will are possible \citep{Pereboom2006LivingWithoutFreeWill}.

Hard incompatibilism seems to be the most logical conclusion of physicalism. Physicalism claims that consciousness is reducible to the insentient substance that we commonly call matter, and everything supervenes on the micro-physics. In the philosophical and psychological discussions of free-will, the ability to make decisions cannot be separated from the agent's consciousness. But when looking at the micro-physical level, all we see is particles and fields blindly following the physical law. Intention, decision, consciousness, all of these are supposed to emerge from this. Then, there should exist an evolutionary explanation of why we developed the sense of free-will: likely because in order to survive, we needed to care about ourselves \citep{Dennett2015ElbowRoomTheVarietiesOfFreeWillWorthWanting}. To care about ourselves, it helps if we behave as if we take ownership of our behavior, if we adopt as our own whatever changes occur in our physical states due to the laws of physics. So the implication of physicalism is most likely that free-will is a useful illusion, whether we're talking about deterministic or indeterministic behavior.
I'd like to add that many physicalists support some versions of free-will, if defined in a way that avoids talking about anything else but the dynamics of the physical stuff, interpreted as due to conscious agency only in a weakly emergent sense \citep{Dennett2015ElbowRoomTheVarietiesOfFreeWillWorthWanting,Carroll2017TheBigPictureOnTheOriginsOfLifeMeaningAndTheUniverseItself,Carroll2021ConsciousnessAndTheLawsOfPhysics}, or where the illusion of freedom may be due to our inability to predict the result of our internal deliberation, as explained for example in \citep{aaronson2013ghostQM}.

If quantum mechanics ensures the existence of genuine indeterminism in nature, how could it be used for free-will? Is the agent supposed to use a deterministic mechanism to decide what choice to make, and then to use the causal opening offered by wavefunction collapse to enforce her decision, determined by the micro-physics, to the physical world? From a physicalist point of view, this seems to make little sense, because the decision itself would be deterministic. And if indeterminism is involved in the decision-making, how is it different from just going along with whatever the result of the wavefunction collapse is, in the exact same way compatibilists go along with the deterministic processes?

Whatever proposals to account for genuine free-will we may make, it seems to me that they cannot be purely physicalistic, that we have to make additional hypotheses that somehow are consistent with the physical laws, but without actually being reducible to them. Since these hypotheses can't be read directly from the micro-physics without requiring an interpretation, a narrative, they should be seen as metaphysical assumptions. Any experimental test that can be reproduced by third parties will report on the behavior of the matter following the physical laws, and only with additional insights can it be interpreted as decision-making of conscious agents. These additional insights have no origin in the uninterpreted reading of the micro-physical events, they come from our internal experiences \citep{Swinburne2013MindBrainAndFreeWill,sep-freewill}.
Without having these experiences of freedom of choice, illusory or not, it would probably never have occurred to us that there is such a thing as free-will, just by examining the structure and dynamics of matter.
Moreover, perhaps only because we have these experiences ourselves, we attribute the existence of similar experiences to other agents that we study, because we see them as not being fundamentally different from ourselves.
So we know from our experiences that we feel that we are agents able to make choices, we interpret these experiences based on metaphysical assumptions, and we overlap these interpretations on the physical laws. But if we try to be authentic physicalists or materialists, if we go all the way down, all remains is blindly interacting particles and fields, so we are forced to deny our own experience and say that free-will, along with consciousness, are some emergent behaviors, even illusions \citep{Dennett2016Illusionism,Frankish2016IllusionismAsATheoryOfConsciousness,sep-eliminative-materialism}.
And indeed, even if we feel that we used our free-will to make a choice, and that we could have done otherwise, in many instances neuroscience tells us that this was not the case \citep{Dennett2015ElbowRoomTheVarietiesOfFreeWillWorthWanting,Sapolsky2023DeterminedLifeWithoutFreeWill}, so at least we know that at least libertarian free-will is often, though not necessarily always, an illusion.
Physics, and natural sciences in general, say very little about free-will and consciousness \emph{per se}, they offer a rigid frame in which we project our metaphysical intuitions and preferences.

More precisely, how are we supposed to determine by experiment, even if we would be able to conduct it at the lowest scales and we collect all details about the micro-physics, if a certain indeterministic event happened by itself, for example due to the wavefunction collapse, or it was caused by the agent as a prime mover as Chisholm proposed?
In fact, Chisholm acknowledges this problem \citep{Chisholm1964HumanFreedomAndTheSelf}:

\begin{fquote}
What, for example, is the difference between A's just happening, and the agents' \emph{causing} A to happen?
We cannot attribute the difference to any event that took place within the agent.
\end{fquote}

His solution is simply an appeal to \emph{tu quoque}:

\begin{fquote}
It is a problem that must be faced by anyone who makes use of the concept of causation at all; and therefore, I would say, it is a problem for everyone but the complete indeterminist.
\end{fquote}

That is, both the libertarian who thinks that the agent really has agency and the compatibilist types of free-will suffer of the same problem: the micro-physics of the agents works well without having to assume that the agents contribute with anything more than just being subsystems in the world. Occam's razor invites the physicalist asked about free-will to reply in the same way Laplace replied to Napoleon when he asked him about God: ``Sire, I had no need of that hypothesis.'' The only difference is that, physicalists or not, we seem to experience free-will, and we can take it as real, as an emergent phenomenon, or as a mere illusion, albeit a useful one, both for motivating individual actions and for building societies in which we can rely on each other \citep{Cashmore2010TheLucretianSwerve}.

So while any discussion about free-will can use the physical laws as a ``skeleton'', the ``flesh'' that can be added on that skeleton consists of metaphysical assumptions. To my knowledge, despite the fact that so many thinkers tried to solve these issues, $60$ years later the situation remains unchanged. And this situation is not improved by the fact that we still don't know if quantum micro-physics really is indeterministic \citep{vonNeumann1955MathFoundationsQM,GhirardiRiminiWeber1986GRWInterpretation} or deterministic \citep{Bohm1952SuggestedInterpretationOfQuantumMechanicsInTermsOfHiddenVariables,Stoica2021PostDeterminedBlockUniverse}, or effectively indeterministic per branch but deterministic for the whole wavefunction \citep{Everett1973TheTheoryOfTheUniversalWaveFunction}.
We also don't know if our decisions use, at least some times, noise or even genuine quantum indeterminism as a resource \citep{vonNeumann1955MathFoundationsQM,Penrose1989EmperorsNewMind,HameroffPenrose2017OrchORUpdatedReview,tegmark2000decoherenceBrain}.

An observation that I think is important is that many discussions about free-will assume from the start that the universe is not aligned with the will of the agent, as if the agent is not an inherent part of the universe. This divisive attitude may betray an explicit or hidden belief in dualism. But what's important is that it informs the attitude we have with respect to free-will. And, since either the universe is causally closed, or, if it's not, and the only input in the causal structure can only be indeterministic, this attitude of antagonism informs the rejection of free-will by invoking physicalism.
On the other hand, why should we see ourselves as in such a big tension with the universe? Aren't we parts of it? Aren't we here, evolved beings who wonder about the mysteries that we've understood so far and those yet unanswered, precisely because the universe brought us into being? In this sense, there should be no conflict between physical laws and freedom, anymore than the skeleton of our bodies bring some rigidity, but without this rigidity we wouldn't be able to walk or to protect our decision-making machines called brains.

For these reasons, while I am extremely interested in free-will, I have to admit that I don't know if we can use physics to prove any position on free-will. And since everything else that can be added to transform the cold equations into a narrative about agents experiencing or not freedom is added on top of what physics says, when speaking from the point of view of physics I feel the need to say that I don't know what free-will is or even whether it exists. I can't give a mathematical definition or a mathematical model of free-will, without having to appeal to an interpretation. Many tried to model free-will and even consciousness, but what I mean here is that I don't know a model that defines or models free-will by itself, without requiring an interpretation in terms of experiences. Let alone a model that is also physically realizable and empirically testable. But since I can't say for sure that this is a limitation of mathematics and physics, I must assume it's my own, and therefore I feel forced to take an epistemically modest position and simply say ``I don't know''.

And yet, even after admitting that we can't read from the structures of mathematical physics whether they realize or not free-will, nor whether and how we can test them empirically, we can try to see what interpretations or definitions of free-will are logically compatible with these structures.
And since we know that micro-physics is quantum, this can be a starting point of exploring these compatibilities.

Ever since the discovery of quantum mechanics, the postulation of the wavefunction collapse was associated with the most basic form of libertarian free-will, which equates indeterminism with free-will \citep{Heisenberg1958PhysicsAndPhilosophy,Wigner1967RemarksOnTheMindBodyProblemWignersFriend,Wilber2001QuantumQuestionsMysticalWritingsOfTheWorldsGreatPhysicists,Stapp2015QuantumMechanicsMindBrainConnection}.
More and more refined views about libertarian free-will developed over time, for example Chisholm's account of libertarian free-will in terms of the agent being a ``prime mover unmoved'', which seems to be echoed by Wheeler when discussing quantum measurements \citep{Wheeler1983LawWithoutLaw}:

\begin{fquote}
To use other language, we are dealing with an elementary act of creation.
\end{fquote}

If this provides an acceptable basis for free-will for those who think that the unpredictability of the results of quantum measurements leave room for free-will, others don't find such interpretations of quantum mechanics compelling for other reasons. A notable example is Wheeler's student Everett, who proposed that all results of quantum measurements exist. After all, if the Schr\"odinger equation is a universal law, why would this law be suspended and replaced with the wavefunction collapse when we do a quantum measurement, as if the measuring device isn't a quantum system itself?
Anyway, I don't intend to address the usual questions about Everett's proposal, \emph{the many-worlds interpretation} (MWI), in its original form or in different variations (my modest answers to these questions can be found in \citealp{Stoica2023TheRelationWavefunction3DSpaceMWILocalBeablesProbabilities,Stoica2024ClassicalManyWorldsInterpretation}). The scope of this article is limited to the possible relations between MWI and free-will. I will use as a main pretext Nicolas Gisin's criticism of free-will in MWI, because it manages to concentrate more lines of attack, some of them novel and profound, in a concise form.

\section{The many-worlds interpretation and free-will}
\label{s:FW-MWI}

As I argued above, some of the perceived tensions between free-will and the physical laws come from seeing individual freedom as ``me against the universe''. In the many-worlds interpretation, this can reach a whole new level, ``me against the multiverse''.
In a recent very entertaining one-page article \citep{Gisin2022MultiversePandemic,Gisin2010LEpidemieDuMultivers}, Nicolas Gisin raises profound questions about free-will in MWI.
He charges MWI on three counts, one for each of the following features it has \citep{Everett1973TheTheoryOfTheUniversalWaveFunction,deWittGraham1973ManyWorldsInterpretationOfQuantumMechanics,Wallace2012TheEmergentMultiverseQuantumTheoryEverettInterpretation,SEP-Vaidman2021MWI}:

\begin{ftab}
\begin{feat}[Determinism]
\label{feat:determinism}
Since its dynamics is given by the Schr\"odinger equation only, MWI is deterministic.
\end{feat}

\begin{feat}[Multiple alternatives]
\label{feat:multiple-result}
Everything that has a nonzero amplitude to happen, happens in some world.
\end{feat}

\begin{feat}[High-level of entanglement]
\label{feat:entanglement}
Everything seems to be entangled with everything else, limiting creativity.
\end{feat}
\end{ftab}

Because of the difficulties to define or to describe them in terms of physical data, mentioned in the Introduction, I don't really know how to define free-will or creativity physically, all I know is my own subjective experience of freedom. This experience may be in fact a compatibilist free-will experienced as libertarian free-will, or it may be an illusion altogether. But in a world made of apparently insentient physical stuff endowed with no intrinsic intentionality, this is our only source of information that there is free-will.
But we don't need this, because if whatever we call free-will corresponds to our experience, its compatibility with MWI should boil down to whether MWI supports human beings having the experience of free-will just like in the interpretations of quantum mechanics (QM) considered compatible with free-will.

\subsection{Determinism in MWI and free-will}
\label{s:FW-MWI-determinism}

\begin{fquest}
\begin{chcount}
\label{count:determinism}
In a deterministic world, we have no freedom.
\end{chcount}
\end{fquest}
\begin{proof}[Reply]
According to \emph{compatibilism}, it is perfectly possible that our will is compatible with a causally closed world.
But this may seem to be a too simplistic semantic trick to avoid the problem, and there is more to be said.

But how can indeterminism allow free-will? How would it help if our decisions are not fully determined by our own present state, but by occasional indeterminism breaking into the causal chain?

\emph{Wouldn't we be more free if we can determine our next decisions based on how we are now, rather than letting them at the mercy of indeterminism?}

Gisin mentions Descartes' solution, that ``mind'' is ontologically distinct from ``matter'', and our will affects the physical by some ``openings'' in its causal chain.

But then, if dualism {\citep{sep-dualism}} is true and accounts for libertarian free-will, the stuff making our mind, our will, should have its own logically consistent laws. And since decision presupposes change, these laws should be those of a dynamical system, deterministic or not.
The ``will-stuff'' or the ``mind-stuff'' (\emph{res cogitans}) should have its own structure and its own dynamics, even if its nature is mental, similar to how the ``matter-stuff'' or the physical-stuff (\emph{res extensa}) has its own laws and dynamics.
Then, if the ``will-stuff'' interacts with the physical-stuff, they form together a larger dynamical system \citep{Stoica2020NegativeWayToSentience}, governed by some laws just like physical systems are. The difference is just that it has a double ontology, ``matter'' and ``mind'', but this is irrelevant for how it follows the laws. So we gained nothing, the ``openings'' in the causal chain are just gates to a larger causal chain, and the questions return.

Another possibility to avoid the problem that indeterminism by itself does not guarantee libertarian free-will is by making an assumption which leads to no physically detectable difference, a metaphysical one, like the one proposed by Chisholm, in terms of the agent being a ``prime mover unmoved''. Suppose that the agent acts like a prime mover unmoved by using the wavefunction collapse, turning it, as Wheeler put it, into ``an elementary act of creation'' {\citep{Wheeler1983LawWithoutLaw}.}

But let us recall the full context of Wheeler's quote:

\begin{fquote}
To use other language, we are dealing with an elementary act of creation.
It reaches into the present from billions of years in the past. It is wrong to think of the past as ``already existing'' in all detail. The ``past'' is theory. The past has no existence except as it is recorded in the present. By deciding what questions our quantum registering equipment shall put in the present we have an undeniable choice in what we have the right to say about the past.
\end{fquote}

We see that Wheeler's \emph{participatory universe} is more than simply using indeterminism as a backdoor in the causal chain of events. In addition to the wavefunction collapse,
\emph{there is another opening in the causal chain: the initial conditions}. What if the initial conditions are not fully specified at the beginning of time, but are gradually determined as more observations and choices are made? As if God left some blank parameters defining the initial conditions of the deterministic universe, to be filled in later by our own choices. Independently, Hoefer makes the case for this ``inside-out freedom'' in classical special relativity, in \citep{Hoefer2002FreedomFromTheInsideOut}.

But if we accept Wheeler's idea of agents that participate to the initial conditions, we can apply it much strongly, to avoid the violations of unitary evolution apparently required by the wavefunction collapse, but also the branching into more worlds. This should be possible too, if our choices \emph{now} can contribute to the initial conditions at the beginning of the universe, as I explained in \citep{Stoica2008ConvergenceAndFreeWill,Stoica2008FlowingWithAFrozenRiver,Stoica2021PostDeterminedBlockUniverse} (also see Scott Aaronson's articles building on this idea in {\citealp{aaronson2013ghostQM}}). If there are yet unspecified values for some degrees of freedom, left blank, they can be adjusted later to lead to a unique outcome for each measurement without having to collapse or branch the wavefunction. These blank parameters may be filled in later, when choosing the measurement settings, so that quantum measurements have definite results without changing the Schr\"odinger equation, without collapse, without adding new variables, and without creating new worlds.

This can be understood as ``superdeterminism'', if we take the view that the initial conditions are already given at the beginning, but in such a way as to ensure the uniqueness of the outcomes of measurements without appealing to collapse or branching. Or it can be understood as ``retrocausality'', if we adopt a four-dimensional block universe view as in relativity, but allow the state of the universe to be specified by constraints distributed at various places and moments in time, that is, not all of them being specified at the Big Bang {\citep{Stoica2021PostDeterminedBlockUniverse}}.

I think the best way to think of them is in terms of sheaves of local solutions that can be extended to global solutions \citep{Stoica2012QMGlobalAndLocalAspectsOfCausalityInQuantumMechanics,Stoica2021PostDeterminedBlockUniverse}.
And, as a bonus, it allows free-will as prime mover unmoved, so a libertarian kind of free-will, but in a deterministic world whose initial conditions are not yet determined {\citep{Stoica2008ConvergenceAndFreeWill}} . This is consistent with Wheeler's interpretation of the delayed choice experiment as making the case for a participatory universe, but in a different way, in which unitary evolution is saved (without resorting to many worlds) {\citep{Stoica13TheTaoOfItAndBitSpringer}}.

More recently, Gisin himself used this ``causal opening'' in the initial conditions of the deterministic laws \citep{Gisin2021IndeterminismInPhysicsClassicalChaosAndBohmianMechanicsAreRealNumbersReallyReal}. His argument is that, since exact real numbers contain infinite information, they can't describe the universe.
He uses the idea of filling in the blanks in the imprecision of numbers, as a way by which potentialities become actualities in a deterministic world. He applies this idea to introduce indeterminism in \emph{Bohmian mechanics} \citep{Bohm1952SuggestedInterpretationOfQuantumMechanicsInTermsOfHiddenVariables,SEP-Goldstein2013BohmianMechanics}. By the way, Bohmian mechanics shares with MWI the Features \ref{feat:determinism} and \ref{feat:entanglement}, and some may say Feature \ref{feat:multiple-result} as well\footnote{Even if Bohmian mechanics proposes that the world as we experience it supervenes on the configurations of point-particles, there is nothing to stop similar worlds to supervene on the wavefunction, which has branches just like MWI. These worlds must contain wavefunction patterns (in the sense of MWI, see \citealp{Wallace2012TheEmergentMultiverseQuantumTheoryEverettInterpretation}) arranged and evolving like agents as well. This is necessary because these waves guide the point-particles in their motion. And, when these ``agent-like patterns'' perform experiments, they are affected by Feature \ref{feat:multiple-result}, resulting in multiple outcomes \emph{in the wavefunction's patterns}, just like in MWI.} \citep{Deutsch1996BMasMWIChronicDenial}, so Gisin's arguments, if correct, should apply to Bohmian mechanics too. Conversely, the arguments proposed here can be applied to the guiding wavefunction in Bohmian mechanics too.

However, what I propose here is an application to MWI of the idea that I originally used for a single-world unitary-only evolution. In MWI, there are also initial conditions, and if they can be chosen to ensure a unique outcome, as in {\citep{Stoica2008SmoothQuantumMechanics}}, this should be possible also in MWI, for those cases when the agent uses quantum measurements within her own brain to make decisions. Therefore, in MWI the agents can use the same loophole of ``delayed initial conditions'' to reduce their future choices, even to a single choice.
Similarly, even Gisin's own proposal for infusing free-will in a deterministic classical world or in Bohmian mechanics can work for MWI as well.

The proposal presented here has the advantages of both libertarian and compatibilist free-will, because the agent is free to act as a prime mover unmoved, but she can do this in a way that also defines ``retroactively'' part of her own past, by filling it the values of the degrees of freedom that remained blank and were not revealed by observations up to the moment when she exercises her free-will. The agent can make a free choice now, in the libertarian sense, and realize that this was what she wanted all along but didn't know it until now.
\end{proof}

\subsection{Multiple alternatives in MWI and free-will}
\label{s:FW-MWI-multiple-alternatives}

\begin{fquest}
\begin{chcount}
\label{count:multiple-result}
If every possibility is realized, all our choices are realized, and we have no freedom.
\end{chcount}
\end{fquest}
\begin{proof}[Reply]
If it seems limiting to be forced by micro-physics to make a particular choice, it should probably feel more limiting to be forced to make all possible choices, even those you don't want to make.

But MWI does not force the agent to make all possible choices she may have. MWI ensures branching when the wavefunction spreads over multiple macro-states (which appear to us as classical, see {\citealp{Wallace2012TheEmergentMultiverseQuantumTheoryEverettInterpretation}}), for example when measurements happen, or when atom or particle decays take place. If the agents can harness quantum measurements or decay within their brains in their decision-making, this still doesn't mean that they would make choices in contradiction with their own preferences or desires.

The argument from my reply to Count \ref{count:determinism} applies to laws which, given the initial conditions, determine a single history, while in MWI multiple histories happen.
But also in MWI not all choices happen. The only possible worlds are those consistent with the initial state and unitary evolution, in the sense that these worlds are represented by state vectors that are not orthogonal to the unitarily evolved initial state. This implies that choices that would result in a total state vector that could not have evolved from the initial state are not available to the agent.

But how restricted is the allowed initial state of the universe? MWI requires special initial conditions, otherwise branching into worlds would not happen only towards the future, and the Born rule would not be consistent with the records of past measurements kept in the present state of the world.
But shouldn't branching be, by definition, only ``towards the future''? What would even mean ``branching towards the past''? It simply means that separated worlds, or even ``parallel'' worlds that didn't arrive from branching, can interfere. Suppose that after a spin $1/2$ measurement along an axis we found that the spin is ``up''. According to MWI, if the spin prior to the measurement was a linear combination of ``up'' and ``down'', there is also a branch in which we found the spin to be ``down''. A repetition of the measurement in the first branch should find again that the spin is ``up''. But if the two branches interfere in the meantime, this will affect the result, and we can obtain that the spin is ``down''. Obtaining the spin ``up'' and then, by measuring it again, obtaining that it is ``down'', would be a violation of the Born rule. We normally exclude the possibility that the branches, once separated, can interfere again, by invoking decoherence. But the mechanism of decoherence works only if the initial conditions of the universe are special, in a similar sense in which they have to be special to ensure that the entropy increases. In MWI (and Bohmian mechanics for that matter), for branching to happen only towards the future, the initial conditions have to satisfy more or less the same constraints as those necessary to ensure Second Law of Thermodynamics, as argued by Wallace  (see Chapter 9 in \citealp{Wallace2012TheEmergentMultiverseQuantumTheoryEverettInterpretation}).

\begin{note}
\label{note:conspiracy}
In fact, the initial conditions turn out to be much stricter than simply being of very low-entropy, as it is often believed. This was shown in {\citep{Stoica2024DoesQuantumMechanicsRequireConspiracy}}.
Let me detail this a bit, since it is a counterintuitive and unexpected result, and it may improve, even if only marginally, the argument.
The state space contains all possible state vectors. This means, it contains state vectors that encode the records of results of measurements as they happened, but it also contains state vectors that encode records of things that didn't happen.
For example, a state vector can contain an agent whose brain contains false memories of herself levitating or turning into a unicorn, but it is not plausible that what's in her memory really happened. Similarly, a state vector that contains false records of repeated spin measurements along the same axis, but so that the subsequent measurements didn't confirm the first result, can exist, but such a history should be vanishingly rare. And yet, state vectors encoding an agent with such false memories or containing records in clear violation of the Born rule are perfectly valid state vectors in the quantum state space. The records can ``lie'' about the past. In this case, there is no past history consistent with these records, the states containing them are the result of interference, just like classical \emph{Boltzmann brains} (spontaneous ``coagulations'' of brains with false memories of their past history) are expected to result from fluctuations {\citep{Eddington1934NewPathwaysInScience_Boltzmann_brains}}.
In {\citep{Stoica2024DoesQuantumMechanicsRequireConspiracy}} it was shown that such situations are rather the rule than the exception: in the state space, the state vectors containing agents with false memories akin to Boltzmann brains vastly outnumber those containing only agents with reliable memories. The state vectors containing records that are inconsistent with the actual past history of the system fill the state space, and only a subspace of infinitely smaller dimension that that of the total space contains reliable records. A quick solution to this puzzle may seem to assume that the universe started in a very special low-entropy state. While this is condition is necessary, it was shown in {\citep{Stoica2024DoesQuantumMechanicsRequireConspiracy}} that it is not sufficient, and a condition much stricter than we expect is needed.
The initial conditions have to be very restricted in a way that takes into account the dynamical law and future records of the events.
And it was shown that these conditions should involve strong correlations between any particle and the rest of the universe, even if only to ensure the possibility to build a well-calibrated measuring device ({\citealp{Stoica2024DoesQuantumMechanicsRequireConspiracy}}, Section 3). And this applies to all major interpretations of quantum mechanics. Normally, when imagining quantum experiments, we take for granted the existence of measuring devices, but their very existence requires very special initial conditions. In any interpretation of quantum mechanics, the initial conditions have to depend of the of the laws and perhaps even of future measurement set-ups. This was thought to be exclusively the problem of ``superdeterministic'' theories, but such correlations are required even by those interpretations of QM that don't rely on violations of Statistical Independence in Bell's sense {\citep{Bell2004SpeakableUnspeakable}}.
\qed
\end{note}

The branching asymmetry discussed by Wallace in Chapter 9 of {\citep{Wallace2012TheEmergentMultiverseQuantumTheoryEverettInterpretation}} and the argument from Note {\ref{note:conspiracy}} give additional reasons why the kind of injection of freedom in the initial conditions by filing the blanks later, when our choices take place, may work for MWI too, allowing our will to limit the alternatives.
If the compatibilization of libertarian free-will with the causal chain by filling in the blanks works for a deterministic world, it should work for MWI too.

But, assuming that in MWI the agents' decision-making can itself be subject to branching, why being restricted to a unique choice would mean more freedom than making all possible choices in different worlds?
\emph{A world in which we can choose only one thing and all the others are forbidden restricts our freedom.}
MWI allows us to follow Yogi Bera's advice,

\begin{fquote}
When you come to a fork in the road, take it.
\end{fquote}

If Alice has to choose between two mutually exclusive options ``coffee'' and ``tea'', and wants them both, MWI allows her to choose both of them, albeit in different worlds.
In the world where Alice's choice is coffee, she experiences doing this by her own free-will, and similarly in the world where her choice is tea. Any attempt to trace in her brain previous indications that she didn't want to make that choice will find none, because her state can't evolve, even with collapse or branching, in a state that contradicts her past history. The wavefunction has two branches,
\begin{ffig}
\begin{equation}
\label{eq:Alice-wants-AB}
\begin{gathered}
a\left|\textnormal{Alice wants coffee}\right\rangle\left|\textnormal{Alice drinks coffee}\right\rangle\\
\text{and} \\
b\left|\textnormal{Alice wants tea}\right\rangle\left|\textnormal{Alice drinks tea}\right\rangle,
\end{gathered}
\end{equation}
\end{ffig}
\noindent where $a,b$ are complex numbers so that $a^2+b^2=1$.
If MWI gives Alice the freedom to make both choices, it gives her more freedom.

For example, if Alice has to choose between mutually exclusive coffee and tea, she can have it both ways in different worlds,
\begin{fcontrol}
\begin{equation}
\label{eq:checkboxbutton-both}
\begin{aligned}
&\text{\checkboxbutton*}\text{ Alice chooses coffee.}\\
&\text{\checkboxbutton*}\text{ Alice chooses tea.}\\
\end{aligned}
\end{equation}
\end{fcontrol}

In {\eqref{eq:checkboxbutton-both}} I used check boxes, which are not mutually exclusive, while the radio buttons used in {\eqref{eq:radiobutton-coffee}} and {\eqref{eq:radiobutton-tea}} are mutually exclusive.

And if Alice wants coffee but not tea, this means that her own disposition when making the choice was to choose coffee, so that $b=0$,
\begin{fcontrol}
\begin{equation}
\label{eq:checkboxbutton-x-coffee}
\begin{aligned}
&\text{\checkboxbutton*}\text{ Alice chooses coffee.}\\
&\text{\textcolor{disabled}{\checkboxbutton}}\text{ \textcolor{disabled}{Alice chooses tea.}}\\
\end{aligned}
\end{equation}
\end{fcontrol}

Similarly, if she wants tea but not coffee, her disposition corresponds to $a=0$,
\begin{fcontrol}
\begin{equation}
\label{eq:checkboxbutton-x-tea}
\begin{aligned}
&\text{\textcolor{disabled}{\checkboxbutton}}\text{ \textcolor{disabled}{Alice chooses coffee.}}\\
&\text{\checkboxbutton*}\text{ Alice chooses tea.}\\
\end{aligned}
\end{equation}
\end{fcontrol}

Alice's preference for coffee can be understood in the compatibilist sense, as being determined by the initial conditions of the universe, but also in the libertarian sense, as Alice choosing now how to fill in the blanks in these initial conditions.
Therefore, Alice can exercise her status of unmoved prime mover to participate to the initial state of the universe, which is far in the past, by making her choice in the present.

At the level of micro-physics, in MWI, this would work in the following way. In {\citep{Stoica2008SmoothQuantumMechanics}} I proposed that the previous system with which the observed system interacted, usually the preparation device, becomes entangled with the observed system. This entangled state is a linear combination of tensor products between eigenstates of the observables and states of the preparation device. When the measurement is completed, the resulting state is such a term of the product, so that the resulting eigenstate of the observed system is accompanied by a state of the preparation device, and an interaction between them taking place in a way that can ensure the conservation laws and even unitarity.
For each of these unitary histories, this interaction with the preparation device is like the spontaneous ``kick'' coming from the environment and affecting the observed particle, ensuring thereby the unitary evolution for that history, as postulated by Schulman in his ``special states'' proposal {\citep{schulman1997timeArrowsAndQuantumMeasurement}}.
(Note that if we don't take into account the past interactions of the observed system, quantum measurements violate the conservation laws, even if we take the measuring device into account {\citep{Burgos1993ConservationLawsQM}}, \citep{Stoica2017TheUniverseRemembersNoWavefunctionCollapse,Stoica2021PostDeterminedBlockUniverse}.)
Later, motivated as well by recovering the conservation laws, Collins and Popescu rediscovered this solution based on entanglement and constructed a more detailed theoretical model {\citep{CollinsPopescu2024ConservationLawsForEveryQuantumMeasurementOutcome}}. 
In the case of MWI, we can use this proposal to ensure a continuity between Alice choosing coffee and her past, or Alice choosing tea and her past, so that the choice results from her own preferences or tendencies. This can accommodate compatibilist free-will, but also libertarian free-will, if Alice can act as a prime mover unmoved to fill in the blank at the moment of choice, so that her past micro-state was already leading to her choice.

It can be objected that, even if MWI allows Alice to make both choices, each version of Alice can enjoy the benefits of only one of these choices in each world. This is consistent with Everett's idea that each branch correlates with her brain being in a classical state \citep{Everett1973TheTheoryOfTheUniversalWaveFunction}.

Can Alice have the experience of enjoying both worlds at once? Can her mind be in such a quantum state? My classical mind writing these words is unable to grasp such a quantum mind, but that's just my classical mind.
Given that any discussion of free-will is anyway deemed to supplement the physics with metaphysical assumptions, we can as well entertain the metaphysical claim that Alice is more than we see in a single branch, that it has a super-mind containing the instances of her classical minds in more branches at once, and maybe unifying them in a higher form of her self. But perhaps such a speculation is too wild and unnecessary, given that most discussions of free-will make minimal metaphysical claims, and even those perhaps with the sole purpose of accounting for our experience of freedom. And it is unnecessary to go that far, given that I already gave a couple of reasons why the many worlds, with their multiple choices for Alice, don't constrain her freedom, but they may in fact fulfill it even more.

However, I'd like to add another argument. Libertarian free-will assumes the existence of possible alternatives from which the agent can choose freely. But here I think lies another conflict with the physical reality: things either exist or don't exist. While we talk about ``possibilities'', we don't know of any example of something that is neither real, nor unreal, but it is possible. Possibility seems to be in a realm between real and unreal, and we have neither a physical understanding of this, nor a mathematical model. From a structural-realist {\citep{sep-structural-realism}} point of view, there is no difference between the structure of a possible world and that of a real world. And yet, we use the idea of possibility often, in the general guise of ``counterfactuals''. We use counterfactuals when discussing probabilities, because probabilities are the ratio between the number of possible favorable events and the total number of possible events, even though only one of these events happened. We use counterfactuals when discussing libertarian free-will, for example Swinburne wrote in {\citep{Swinburne2013MindBrainAndFreeWill}}, p. 203:

\begin{fquote}
It is natural to suppose that there follows from someone having free will in my sense a principle called `the principle of alternative possibilities' (PAP) that:

A does \emph{x} freely only if he could have not done \emph{x} (i.e. could have refrained from doing \emph{x}).
\end{fquote}

Counterfactuals were used by David Hume to define causality, by stating that an event \emph{x} causes another event \emph{y} if and only if, without \emph{x}, \emph{y} would not exist {\citep{DavidHume2000ATreatiseOfHumanNature}}.

Counterfactuals were also used to reject the \emph{triviality argument} against the \emph{computational theory of mind}. According to Putnam {\citep{Putnam1988RepresentationAndReality}} and Searle {\citep{Searle1990IsTheBrainADigitalComputer}}, any physical system can be interpreted as implementing any computation. One of the attempts to refute the triviality argument is that a physical system implements a computation only if it is able to compute for alternative inputs \citep{Chalmers1996DoesARockImplementEveryFiniteStateAutomaton,sep-computational-mind}.

But how can possible alternatives make sense, if something either exists or it doesn't and there is no middle way? And if counterfactuals as possibilities make no physical sense, how can then libertarian free-will make sense?

David Hume, an important proponent of compatibilism, also thought that everything about the world is determined by the configuration of matter and the distribution of particular events, by facts about objects in space and time, and that there is no reason to infer from this the existence of causal relations or of fundamental laws connecting these events. This position, named \emph{Humean supervenience}, deems as unnecessary and unwarranted not only the causal relations and the physical laws as abstract entities connecting the events, but also the possibilities, the counterfactuals, as pointed out for example by Loewer {\citep{Loewer1996HumeanSupervenience}}. This became very clear in David Lewis's \emph{modal realism}, according to which the ``possible worlds'' should be understood as being as real as the actual world \citep{Lewis1973Counterfactuals,Lewis1986OnThePluralityOfWorlds,sep-possible-worlds}.
 Then, there is no ground for libertarian free-will based on alternative possibilities, unless each possible choice is realized in some existing world. But MWI ensures the existence of other worlds, remaining in the game as a viable physical basis for free-will, and maybe as the only such basis.
\end{proof}

\subsection{Entanglement in MWI and free-will}
\label{s:FW-MWI-entanglement}

\begin{fquest}
\begin{chcount}
\label{count:entanglement}
``When everything is entangled with everything else, in one big monstrous piece, there is no room left for creativity'' \citep{Gisin2022MultiversePandemic}.
\end{chcount}
\end{fquest}
\begin{proof}[Reply]
It is said that Leonardo da Vinci worked on the \emph{Mona Lisa} between 1503 and 1517. An artist tries numerous versions, explores numerous potential worlds in a single world.
MWI may allow different versions of da Vinci, with different versions of the Mona Lisa, including the one we know. 
If our own history, with a particular version of the Mona Lisa, involves creativity, how would the same history lack creativity in MWI, just because multiple other variations happen? If ``everything happens'' in MWI, how could creativity not happen?

Could it be true that in MWI the histories in which Shakespeare produced randomly both great and bad literature overwhelmingly dominate the multiverse?
This, and even worse, should be the case if all initial conditions would be available.
If MWI gives the same probabilities as standard QM, Shakespeare should create consistently great or consistently bad literature in most histories.

But, as explained in \S{\ref{s:FW-MWI-multiple-alternatives}}, this is not the case: the initial conditions, even for MWI, have to be severely constrained.
Leonardo da Vinci and Shakespeare didn't create their works randomly, by making arbitrary choices. They carefully educated and trained themselves for a long time.
Their precision shielded them both from noise and uncontrollable indeterminism, and from entanglement.
And if they did indeed fill in the blanks in the initial conditions of the universe, they could say along with the Areopagite, who wrote this a millennium before Michelangelo could have said it \citep{AreopagiteWatts1994TheologiaMystica},

\begin{fquote}
For this is not unlike the art of those who hew out a life-like image (from stone), removing from around it all which impedes clear vision of the latent form, showing its true and hidden beauty solely by taking away
\end{fquote}
\noindent and pruning by this the undesired branches of the wavefunction.

But maybe entanglement introduces additional constraints that should be considered.
So how would entanglement limit creativity?

While any interpretation of QM contains entanglement, MWI contains much more, because it is based on decoherence. In each world, the measured system is separated from the environment, so each world has the same amount of entanglement as in standard QM. But in the total wavefunction, containing the many worlds, the observed system is entangled with the environment.
Every time new worlds are created, new entanglement is produced. Standard QM avoids this by collapsing the wavefunction at the end of each measurement, so that in the end the observed degrees of freedom are not entangled with the environment.
But in MWI, more entanglement is produced with each new measurement.
The same amount of entanglement is present in Bohmian mechanics, which requires the same branching structure as MWI, otherwise the ``empty branches'' will interfere with the one correlated with the Bohmian positions, making the macroscopic objects unstable and violating the Born rule.

In each world, the entanglement is exactly how it has to be in standard QM.
And what happens in one world is not affected by the other worlds, unless previously separated worlds interfere again, which would be a bigger problem for MWI than too much entanglement.

Returning to the ``opening'' in the causal chain that may be needed for free-will, in a deterministic world, even with many-worlds, the more ways to fill in the blanks in the initial conditions, the more possibilities of freedom exist. And entanglement only adds more possibilities, more parameters with more blanks to be filled in.

If we evolved to use the resources of pseudo-indeterminism like noise, and those of genuine indeterminism like that resulting from quantum measurements, since entanglement is also a resource {\citep{Wootters1998QuantumEntanglementAsAQuantifiableResource}}, we may as well have evolved to use entanglement too. In fact, entanglement is the main resource on which modern quantum technologies are based {\citep{Chitambar2019QuantumResourceTheories}}.
Whatever abilities we developed during our evolution, including what we call free-will or creativity, are not due to the subsystems alone against the environment, but to the complex interplay between them. \emph{Are these abilities properties of us as subsystems, or of the whole?}
If quantum mechanics, in particular its many-worlds interpretation {\citep{Pas2024TheOneHowAnAncientIdeaHoldsTheFutureOfPhysics}}, but also Bohmian mechanics {\citep{BohmHiley1993UndividedUniverse}} taught us something, is that the universe is an undivided whole. Therefore,

\emph{We are interconnected with the rest of the universe, and maybe these connections enchain us, or maybe we evolved to use them to affect the world.}

If, during the evolution of life, this resource of entanglement could be used for the survival of the individuals and of the species, it was used. But at any rate, if it was not used, its existence does no harm, and adds no limitation. If creativity is possible in Standard QM, and if any history possible in Standard QM is also possible in MWI, creativity is possible in MWI as well.
\end{proof}

\section{Conclusions}
\label{s:conclusions}

I tried to take a rather prudent position, by separating the physical from the metaphysical assumptions involved in what is understood by free-will.
In this spirit of prudence and epistemic modesty, I prefer to say that I don't know what free-will is, other than my own subjective experience of freedom, which informs, as in the case of anyone else, my views on this issue.
But even with this provision, it is possible to analyze what types of free-will are consistent with MWI.

I want to emphasize again that libertarian free-will depends on whether the brain is able to use quantum indeterminism or the branching resulting from decoherence, as a resource in the decision-making processes.
Compatibilist free-will depends on whether the brain is able to realize a chain of commands able to lead to the desired outcomes, even if noise or genuine indeterminism interfere with its actions.
The article should be understood with these provisions, and with the provision that the physical structures don't have more to bring to the table than whether their behavior is deterministic or not, and retrocausal or not.

While I defended the compatibility of free-will with MWI, I only expressed some personal views about logical possibilities, using as a starting point Nicolas Gisin's excellently written, very concise article. In my opinion, there is much agreement between my views and his with respect to getting a chance to have free-will even in an apparantely deterministic but incompletely specified universe. The disagreements come with respect to MWI, about which I argue that it supports free-will.

In making the case for the compatibility of free-will in MWI, but also when expressing cautions, I brought some arguments that, to my best understanding, are novel:

\begin{enumerate}
	\item 
	Libertarian free-will based on dualism should, from a structural-realist and physicalist point of view, lead to a dynamical system composed of a matter-system and a mind-system that interact, and it has the same problems as the purely physicalist or materialist systems regarding the possibility of free-will (\S\ref{s:FW-MWI-determinism}).
	\item 
	Libertarian-style free-will is possible even in a deterministic world, and even in deterministic many-worlds, if not all parameters were fixed at the Big Bang but can be fixed later by agents acting as prime movers unmoved (\S\ref{s:FW-MWI-determinism}).
	\item 
	This combines the libertarian and compatibilist types of free-will in a single type of free-will that has both their advantages (\S\ref{s:FW-MWI-determinism}).
	\item 
	The ability to make multiple choices at once in a non-exclusive way gives us not less, but rather more freedom (\S\ref{s:FW-MWI-multiple-alternatives}).
	\item
	If classical agents can be supported by branches in the wavefunction, maybe quantum agents, supported on the full wavefunction and supervening over more branches, are possible as well, but we only experience a ``classical component'' (\S\ref{s:FW-MWI-multiple-alternatives}).
	\item 
	The standard flavors of libertarian free-will require non-physical possibilities to have a sort of reality that is not quite real, but also not unreal. I side with David Lewis that this requires possible worlds to be real \citep{Lewis1973Counterfactuals,Lewis1986OnThePluralityOfWorlds}, and I think that MWI provides a physical basis for this  (\S\ref{s:FW-MWI-multiple-alternatives}).
	\item
	The huge entanglement present in MWI, rather than limiting our freedom, could be a resource, a source of more blanks to be filled-in during our choices (\S\ref{s:FW-MWI-entanglement}).
\end{enumerate}

I conclude that, in my opinion, MWI is compatible with free-will just like other theories and interpretations of quantum mechanics are, with its own flavors that can be argued to bring some advantages.
I'll let the readers use their free-will to decide if these arguments are as plausible as the usual arguments for other versions of free-will, and if they bring something both new and useful to the vast and sophisticated discussion about free-will that is going on since ancient times.

\addcontentsline{toc}{section}{\refname}


\end{document}